\input harvmac
\input epsf
\noblackbox

%%% Paragraphs
\newcount\figno

\figno=0
\def\fig#1#2#3{
\par\begingroup\parindent=0pt\leftskip=1cm\rightskip=1cm\parindent=0pt
\baselineskip=11pt
\global\advance\figno by 1
\midinsert
\epsfxsize=#3
\centerline{\epsfbox{#2}}
\vskip 12pt
\centerline{{\bf Figure \the\figno} #1}\par
\endinsert\endgroup\par}
\def\figlabel#1{\xdef#1{\the\figno}}
\def\pano{\par\noindent}
\def\smno{\smallskip\noindent}
\def\meno{\medskip\noindent}
\def\bigno{\bigskip\noindent}

%%% special math symbols
\font\cmss=cmss10
\font\cmsss=cmss10 at 7pt

\def\rlx{\relax\leavevmode}
\def\inbar{\vrule height1.5ex width.4pt depth0pt}
\def\IC{\relax\,\hbox{$\inbar\kern-.3em{\rm C}$}}
\def\IR{\relax{\rm I\kern-.18em R}}
\def\IN{\relax{\rm I\kern-.18em N}}
\def\IP{\relax{\rm I\kern-.18em P}}
\def\ZZ{\rlx\leavevmode\ifmmode\mathchoice{\hbox{\cmss Z\kern-.4em Z}}
 {\hbox{\cmss Z\kern-.4em Z}}{\lower.9pt\hbox{\cmsss Z\kern-.36em Z}}
 {\lower1.2pt\hbox{\cmsss Z\kern-.36em Z}}\else{\cmss Z\kern-.4em Z}\fi}

%%% misc.
\def\narrowplus{\kern -.04truein + \kern -.03truein}
\def\narrowminus{- \kern -.04truein}
\def\narrowminussub{\kern -.02truein - \kern -.01truein}

\def\o#1{\overline{#1}}

\def\th#1#2{\vartheta\bigl[{\textstyle{  #1 \atop #2}} \bigr] }

%%% further macros

%%% References

\lref\KolbVQ{ E.~W.~Kolb and M.~S.~Turner,
{\it The Early Universe},
Redwood City, USA: Addison-Wesley (1990)
%547 p. (Frontiers in physics, 69)
. }

\lref\rlindebook{ A.~D.~Linde,
{\it
Particle Physics and Inflationary Cosmology},
Harwood, Chur, Switzerland (1990). }

\lref\guth{A.~H.~Guth,
{\it The Inflationary Universe: A possible solution to the Horizon and Flatness
Problems}, Phys. Rev. {\bf D23} (1981) 347.
%%CITATION = PHRVA,D23,347;%%
}

\lref\linde{A.~D.~Linde, {\it A new Inflationary Universe Scenario: A possible
solution of the Horizon, Flatness, Homogeneity, Isotropy and Primordial
Monopole Problems}, Phys. Lett. {\bf B108} (1982) 389.
%%CITATION = PHLTA,B108,389;%%
}

\lref\rbailin{D.~Bailin, G.~V.~Kraniotis and A.~Love,
{\it Standard-like Models from Intersecting D4-branes},
hep-th/0108131.
%%CITATION = HEP-TH 0108131;%%
}

\lref\albrecht{A.~Albrecht and P.~J.~Steinhardt, {\it Cosmology
for Grand Unified Theories with Radiatively Induced Symmetry
Breaking}, Phys. Rev. Lett. {\bf 48} (1982) 1220.
%%CITATION = PRLTA,48,1220;%%
}

\lref\KhouryWF{
J.~Khoury, B.~A.~Ovrut, P.~J.~Steinhardt and N.~Turok,
{\it The ekpyrotic universe:
Colliding branes and the origin of the hot big  bang},
Phys. Rev. {\bf D64} (2001) 123522,
hep-th/0105199.
%%CITATION = HEP-TH 0105199;%%
}

\lref\hybridb{A.~D.~Linde, {\it Hybrid inflation}, Phys. Rev. {\bf D49}
(1994) 748, astro-ph/9307002.
%%CITATION = ASTRO-PH 9307002;%%
}

\lref\hybrid{A.~D.~Linde, {\it Axions in Inflationary Cosmology}, Phys. Lett.
{\bf B259} (1991) 38.
%%CITATION = PHLTA,B259,38;%%
}

\lref\rangles{M.~Berkooz, M.~R.~Douglas and R.~G.~Leigh, {\it Branes
Intersecting at Angles}, Nucl. Phys. {\bf B480} (1996) 265, hep-th/9606139.
%%CITATION = HEP-TH 9606139;%%
}

\lref\berlin{R. Blumenhagen, B. K\"ors and D. L\"ust,
{\it Moduli Stabilization for Intersecting Brane Worlds in Type 0$\, '$
String Theory}, hep-th/0202024.
%%CITATION = HEP-TH 0202024;%%
}

\lref\radd{N.~Arkani-Hamed, S.~Dimopoulos, and G.~Dvali, {\it The Hierarchy
Problem and New Dimensions at a Millimeter}, Phys. Lett. {\bf B429} (1998)
263, hep-ph/9803315.
%%CITATION = HEP-PH 9803315;%%
}

\lref\raadd{I.~Antoniadis, N.~Arkani-Hamed, S.~Dimopoulos, and G.~Dvali, {\it
New Dimensions at a Millimeter to a Fermi and Superstrings at a TeV},
Phys. Lett. {\bf B436} (1998) 257, hep-ph/9804398.
%%CITATION = HEP-PH 9804398;%%
}

%\RabadanMT
\lref\rrab{
R.~Rabadan,
{\it Branes at angles, torons, stability and supersymmetry},
Nucl.\ Phys.\ B {\bf 620} (2002) 152,
hep-th/0107036.
%%CITATION = HEP-TH 0107036;%%
}

\lref\rfhs{S.~F\"orste, G.~Honecker and R.~Schreyer, {\it Supersymmetric
$\ZZ_N \times \ZZ_M$ Orientifolds in 4-D with D-branes at Angles},
Nucl. Phys. {\bf B593} (2001) 127, hep-th/0008250.
%%CITATION = HEP-TH 0008250;%%
}

\lref\rbonna{S.~F\"orste, G.~Honecker and R.~Schreyer, {\it
Orientifolds with Branes at Angles}, JHEP {\bf 0106} (2001) 004,
hep-th/0105208.
%%CITATION = HEP-TH 0105208;%%
}

\lref\rbgklnon{R.~Blumenhagen, L.~G\"orlich, B.~K\"ors and D.~L\"ust,
{\it Noncommutative Compactifications of Type I Strings on Tori with Magnetic
Background Flux}, JHEP {\bf 0010} (2000) 006, hep-th/0007024.
%%CITATION = HEP-TH 0007024;%%
}

\lref\rbgklmag{R.~Blumenhagen, L.~G\"orlich, B.~K\"ors and D.~L\"ust,
{\it Magnetic Flux in Toroidal Type I Compactification}, Fortsch. Phys. 49
(2001) 591, hep-th/0010198.
%%CITATION = HEP-TH 0010198;%%
}

\lref\rbkl{R.~Blumenhagen, B.~K\"ors and D.~L\"ust,
{\it Type I Strings with $F$ and $B$-Flux}, JHEP {\bf 0102} (2001) 030,
hep-th/0012156.
%%CITATION = HEP-TH 0012156;%%
}

\lref\rbgka{R.~Blumenhagen, L.~G\"orlich and B.~K\"ors,
{\it Supersymmetric Orientifolds in 6D with D-Branes at Angles},
Nucl. Phys. {\bf B569} (2000) 209, hep-th/9908130.
%%CITATION = HEP-TH 9908130;%%
}

\lref\rbgkb{R.~Blumenhagen, L.~G\"orlich and B.~K\"ors,
{\it A New Class of Supersymmetric Orientifolds with D-Branes at
Angles}, hep-th/0002146.
%%CITATION = HEP-TH 0002146;%%
}

\lref\rbgkc{R.~Blumenhagen, L.~G\"orlich and B.~K\"ors, {\it
Supersymmetric 4D Orientifolds of Type IIA with D6-branes at Angles},
JHEP {\bf 0001} (2000) 040, hep-th/9912204.
%%CITATION = HEP-TH 9912204;%%
}

\lref\ras{C.~Angelantonj, A.~Sagnotti, {\it Type I
Vacua and Brane Transmutation}, hep-th/0010279.
%%CITATION = HEP-TH 0010279;%%
}

\lref\raads{C.~Angelantonj, I.~Antoniadis, E.~Dudas, A.~Sagnotti, {\it Type I
Strings on Magnetized Orbifolds and Brane Transmutation},
Phys. Lett. {\bf B489} (2000) 223, hep-th/0007090.
%%CITATION = HEP-TH 0007090;%%
}

\lref\rbbh{R.~Blumenhagen, V.~Braun, and R.~Helling, {\it Bound States of
D$(2p)$-D0 Systems and Supersymmetric $p$ Cycles}, Phys. Lett. {\bf B510}
(2001) 311, hep-th/0012157.
%%CITATION = HEP-TH 0012157;%%
}

\lref\riba{L.~E.~Ibanez,
{\it Standard Model Engineering with Intersecting Branes},
hep-th/0109082.
%%CITATION = HEP-PH 0109082;%%
}

\lref\rafiruph{G.~Aldazabal, S.~Franco, L.~E.~Ibanez, R.~Rabadan, A.~M.~Uranga,
{\it Intersecting Brane Worlds}, JHEP {\bf 0102} (2001) 047, hep-ph/0011132.
%%CITATION = HEP-PH 0011132;%%
}

\lref\rafiru{G.~Aldazabal, S.~Franco, L.~E.~Ibanez, R.~Rabadan, A.~M.~Uranga,
{\it $D=4$ Chiral String Compactifications from Intersecting Branes},
hep-th/0011073.
%%CITATION = HEP-TH 0011073;%%
}

\lref\rimr{L.~E.~Ibanez, F.~Marchesano, R.~Rabadan, {\it Getting just the
Standard Model at Intersecting Branes}, hep-th/0105155.
%%CITATION = HEP-TH 0105155;%%
}

\lref\rcvetica{M.~Cvetic, G.~Shiu and A.~M.~Uranga,  {\it Three-Family
Supersymmetric Standard-like Models from Intersecting Brane Worlds}
Phys. Rev. Lett. {\bf 87} (2001) 201801,  hep-th/0107143.
%%CITATION = HEP-TH 0107143;%%
}

\lref\rcveticb{M.~Cvetic, G.~Shiu and  A.~M.~Uranga,  {\it
Chiral Four-Dimensional N=1 Supersymmetric Type IIA Orientifolds from
Intersecting D6-Branes}, Nucl. Phys. {\bf B615} (2001) 3, hep-th/0107166.
%%CITATION = HEP-TH 0107166;%%
}

\lref\rott{R.~Blumenhagen, B.~K\"ors, D.~L\"ust and T.~Ott, {\it
The Standard Model from Stable Intersecting Brane World Orbifolds},
Nucl. Phys. {\bf B616} (2001) 3, hep-th/0107138.
%%CITATION = HEP-TH 0107138;%%
}

\lref\rottb{R.~Blumenhagen, B.~K\"ors, D.~L\"ust and T.~Ott, {\it
Intersecting Brane Worlds on Tori and Orbifolds}, hep-th/0112015.
%%CITATION = HEP-TH 0112015;%%
}

\lref\rbonnb{G.~Honecker, {\it Intersecting Brane World Models from
D8-branes on $(T^2 \times T^4/\ZZ_3)/\Omega R_1$ Type IIA Orientifolds},
JHEP {\bf 0201} (2002) 025, hep-th/0201037.
%%CITATION = HEP-TH 0201037;%%
}

\lref\rqsusy{D.~Cremades, L.~E.~Ibanez and F.~Marchesano, {\it
SUSY Quivers, Intersecting Branes and the Modest Hierarchy Problem},
hep-th/0201205.
%%CITATION = HEP-TH 0201205;%%
}

\lref\rbachas{C.~Bachas, {\it A Way to Break Supersymmetry}, hep-th/9503030.
%%CITATION = HEP-TH 9503030;%%
}

%%   inflation

\lref\rquea{C.~P.~Burgess, M.~Majumdar, D.~Nolte, F.~Quevedo, G.~Rajesh
and R.-J.~Zhang, {\it The Inflationary Brane-Antibrane Universe},
JHEP {\bf 0107} (2001) 047, hep-th/0105204.
%%CITATION = HEP-TH 0105204;%%
}

\lref\ralex{S.~H.~Alexander, {\it Inflation from $D-\bar{D}$ Brane
Annihilation}, Phys. Rev. {\bf D65} (2002) 023507,  hep-th/0105032.
%%CITATION = HEP-TH 0105032;%%
}

\lref\rshafi{G.~Dvali, Q.~Shafi and S.~Solganik, {\it D-brane Inflation},
hep-th/0105203.
%%CITATION = HEP-TH 0105203;%%
}

\lref\rshafib{B.-S.~Kyae and Q.~Shafi, {\it Branes and Inflationary Cosmology},
Phys. Lett. {\bf B526} (2002) 379, hep-ph/0111101. 
%%CITATION = HEP-PH 0111101;%%
}

\lref\rrabadan{J.~Garcia-Bellido, R.~Rabadan, F.~Zamora, {\it
Inflationary Scenarios from Branes at Angles},
JHEP {\bf 01} (2002) 036, hep-th/0112147.
%%CITATION = HEP-TH 0112147;%%
}

\lref\rkallosh{C.~Herdeiro, S.~Hirano and  R.~Kallosh, {\it
String Theory and Hybrid Inflation/Acceleration},
JHEP {\bf 0112} (2001) 027, hep-th/0110271.
%%CITATION = HEP-TH 0110271;%%
}

\lref\rtye{G.~Dvali and S.-H.~H.~Tye, {\it Brane Inflation},
Phys. Lett. {\bf B450} (1999) 72, hep-ph/9812483.
%%CITATION = HEP-PH 9812483;%%
}

\lref\rtyeb{G.~Shiu and  S.-H.~H.~Tye, {\it Some Aspects of Brane Inflation},
Phys. Lett. {\bf B516} (2001) 421, hep-th/0106274.
%%CITATION = HEP-TH 0106274;%%
}

\lref\rqueb{C.~P.~Burgess, P.~Martineau, F.~Quevedo, G.~Rajesh
      and R.-J.~Zhang, {\it Brane-Antibrane Inflation in Orbifold and
      Orientifold Models}, hep-th/0111025.
%%CITATION = HEP-TH 0111025;%%
}

\lref\SenSM{
A.~Sen,
{\it Tachyon condensation on the brane antibrane system},
JHEP {\bf 9808} (1998) 012,
hep-th/9805170.
%%CITATION = HEP-TH 9805170;%%
}

%\SenII
\lref\SenII{
A.~Sen,
{\it Stable non-BPS bound states of BPS D-branes},
JHEP {\bf 9808} (1998) 010,
hep-th/9805019.
%%CITATION = HEP-TH 9805019;%%
}

%\MukhanovXT
\lref\MukhanovXT{
V.~F.~Mukhanov and G.~V.~Chibisov,
{\it Quantum Fluctuation And 'Nonsingular' Universe. (In Russian)},
JETP Lett.\  {\bf 33} (1981) 532.
%%CITATION = JTPLA,33,532;%%
}

%%% Title page
\Title{\vbox{
 \hbox{HU--EP-02/09}
 \hbox{SPIN-02/05}
 \hbox{ITP-UU-02/04}
 \hbox{hep-th/0202124}}}
{\vbox{\centerline{Hybrid Inflation in Intersecting Brane Worlds}
}}
\centerline{Ralph Blumenhagen{$^1$}, Boris K\"ors{$^2$}, Dieter L\"ust{$^1$},
and Tassilo Ott{$^1$}}

\bigskip
\centerline{$^1$ {\it Humboldt-Universit\"at zu Berlin, Institut f\"ur
Physik,}}
\centerline{\it Invalidenstrasse 110, 10115 Berlin, Germany}
\centerline{\tt e-mail:
blumenha, luest, ott@physik.hu-berlin.de}
\medskip
\centerline{$^2$ {\it Spinoza Institute, Utrecht University,}}
\centerline{\it Utrecht, The Netherlands}
\centerline{\tt email: kors@phys.uu.nl}
\bigskip
\centerline{\bf Abstract}
\noindent
Non-supersymmetric brane world scenarios in string theory display
perturbative instabilities that usually involve run-away potentials for
scalar moduli fields.
We investigate in the framework of intersecting brane
worlds whether the leading order scalar potential
for the closed string moduli allows to satisfy the slow-rolling
conditions required for applications in inflationary
cosmology.
Adopting a particular choice of basis in field space
and assuming mechanisms to
stabilize some of the scalars, we find that slow-rolling conditions
can be met very generically.
In intersecting brane worlds inflation can end nearly instantaneously
like in the hybrid inflation scenario due to the
appearance of open string tachyons localized at the intersection
of two branes, which signal a corresponding phase transition
in the gauge theory via the condensation of a Higgs field.

\bigskip

\Date{02/2002}
%%% text
\newsec{Introduction}
\smno Recently, there have been various attempts to realize the
inflationary scenario \refs{\guth\linde-\albrecht} of cosmology
within string theory \refs{\rtye\ralex\rshafi\rquea\rtyeb
\rkallosh\rshafib\rrabadan-\rqueb} (an alternative to inflation within
M-theory was proposed in \KhouryWF). In order to do so, it appears
natural to work in a non-superymmetric string vacuum from the
beginning, as such string models provide non-trivial potentials
for the scalar fields of the effective theory. In particular
non-symmetric (non-BPS) D-brane configurations offer a promising
avenue for inflation, as the displacement of D-branes in the extra
dimensions results in a non-vanishing vacuum energy which induces
an inflationary phase in the three uncompactified dimensions. More
concretely, the displacement parameters of the D-branes in the
internal space are identified with the vacuum expectation values
of some Higgs scalar fields in the effective four-dimensional
theory. So far, it remains a major challenge to find explicit
constructions that lead to realistic potentials with suitable
properties for interpreting one of the scalars as the inflaton
field whose vacuum expectation value gives rise to an effective
cosmological constant which then drives the inflationary expansion
of the early universe. The ultimate goal would be to combine
inflation with realistic particle physics as given by the Standard
Model or GUT theories.

A class of phenomenologically interesting string models in this respect is given by so-called intersecting brane worlds
\refs{\rbgkc\rbgklnon\raads\rbgklmag\ras\rafiru\rafiruph\rbkl\rimr\rbonna\rott\rottb\rrab
\rcvetica\rcveticb\rbonnb\rbailin\riba-\berlin}.
As in common brane world scenarios
the gauge fields are confined to topological
defects, D-branes, which spread a four-dimensional Minkowski space and extend
into the internal compact space of total dimension six.
In the present setting, 6+1 dimensional
D6-branes wrap 3-cycles of the internal geometry and intersect each other in certain patterns which determine
the breaking of supersymmetry as well as the low energy spectrum of
chiral fermions \refs{\rbachas,\rangles}.
The absence of gauge and gravitational anomalies is guaranteed by the stringy RR-tadpole cancellation conditions.
An important feature is the fact that the gravitational sector in the bulk
of space-time still can be  supersymmetric at tree level, while it may be
broken in the gauge theory on the branes.
Non-supersymmetric and supersymmetric models
of this kind have been studied in the literature. It has turned out that one can
find examples of non-supersymmetric compactifications that come very
close to the Standard Model of particle physics.
In particular it was possible to construct models which contain
precisely the gauge fields and the chiral fermions of the Standard Model,
the only difference showing up in the Higgs sector.

But these constructions are plagued by the same problems as all
non-supersymmetric string vacua or, respectively, non-BPS brane
configurations that have been found so far, namely, the appearance
of instabilities of scalar fields due to uncanceled massless NSNS
tadpoles. In \rott\ the leading order perturbative potential for
the closed string moduli fields has been investigated with the
result that it usually implies run-away instabilities for some of
the scalars that were pushed to a degenerate limit. While in
\refs{\rott,\berlin} strategies to avoid such dangerous behavior
were advocated, one may also contemplate to make use of these
potentials by reinterpreting them on cosmological scales.

This is the purpose of the present paper: We investigate whether
an inflationary scenario can be realized in an intersecting brane
world. In doing so, we study the scalar potentials that arise in
these models at the leading and next-to-leading order in string
perturbation theory and then look for suitable candidates for the
inflaton field. The typical inflationary scenario requires a
scalar field $\psi$ whose contribution to the overall energy is
dominated by its potential term and thus works as an effective
cosmological constant roughly given by its vacuum expectation
value. A key criterion to decide if any scalar field is
appropriate is the issue of slow-rolling. During its evolution the
perturbations must be very tiny in order to fit the bounds set by
the highly homogeneous CMB data. This is conveniently rephrased in
terms of the potential $V(\psi)$, which must obey \eqn\slowroll{
\epsilon = {M_{pl}^2 \over 2} \left( {V'(\psi) \over V(\psi)}
\right)^2 \ll 1, \quad \eta = M_{pl}^2 {V''(\psi) \over V(\psi)}
\ll 1. } These two conditions (see for instance
\refs{\KolbVQ,\rlindebook}) of course need to be complemented by
many more checks like the possibility to get 60 e-folding and a
realistic fluctuation spectrum. Therefore it only serves as a
minimal requirement  needed for $\psi$ to be called an inflaton
candidate.

Moreover, in inflationary cosmology one needs a
mechanism to end inflation ``gracefully''. In the context of
intersecting brane worlds, a quite natural exit from inflation is
provided  by the appearance of open string tachyons after some
evolution of closed string moduli. This exit fits into the pattern
of a hybrid inflationary model \refs{\hybrid,\hybridb}
effectively described by a term
\eqn\hybrd{ V^{\rm YM} (\psi,H)\sim( M(\psi)^2 - {1\over 4} H^2)^2
}
in the scalar potential of the gauge theory sector. It
combines the merits of both chaotic inflation and spontaneous
gauge symmetry breaking. The slowly evolving inflaton
field $\psi$ affects the mass $M(\psi)^2$ of a second scalar
field, the Higgs field $H$. When $M(\psi)^2$ becomes negative, a phase
transition occurs and inflation ends immediately. In string theory
open string tachyons may take the role of this scalar signaling also
a phase transition in string theory \refs{\SenII,\SenSM}:
the condensation
of higher dimensional D-branes into lower dimensional ones, respectively
the condensation of two intersecting D-branes into a single one
wrapping a non-trivial supersymmetric 3-cycle \rbbh.
Intriguingly, these tachyonic scalars are well suited to serve
as Higgs fields and drive the gauge theoretic spontaneous symmetry
breaking mechanisms \refs{\rbachas,\rbgklnon,\rafiruph,\rimr}. This could
naturally link the exit from inflation to some phase transition in
the gauge sector, possibly even the electroweak phase transition
itself.

Coming back to the main point, we want to identify the inflaton $\psi$
with some scalar field corresponding to a geometric modulus of the
model. Conceptually, there are two different classes of such
scalars. The first class contains the closed string moduli, some of
them parameterize the shape and size of the geometric, six-dimensional
gravitational background space.
In our cases, this space is given by either a six-dimensional torus or
a toroidal orbifold. In particular, there are the K\"ahler moduli for the size of the
space, the complex structure moduli for its shape, as well as
twisted moduli localized at the singularities (fixed points) of
the orbifolds. The second class contains the positions of the
D-branes on the internal space and the Wilson lines of gauge
fields along the branes. They are open string moduli.

From earlier work \refs{\rquea,\rrabadan}
one expects that the open string moduli
could satisfy slow-rolling properties if one
makes the severe simplification to assume
that the closed string moduli are frozen.
Geometrically speaking, this means that if the background space is fixed
and no backreaction
on the presence of the branes takes place, their motion along this
space can be very slow for a certain time. After this time, they start
approaching
each other faster and at a critical distance  a tachyon appears to signal
their condensation.
Since the dynamics of the entire setting is determined by the
fastest rolling field, this assumption
implies that the closed string moduli have to roll slower than
the open string moduli.
Otherwise, the space could for instance shrink very quickly and bring the two branes
within their critical distance  much faster than originally estimated
from the simplified analysis with frozen volume.
The problem of the simultaneous slow-rolling properties of all
scalar fields involved will be addressed in the present paper and
solutions which allow slow-rolling closed string fields will be proposed.

The paper is organized as follows.
In section 2 we give a very short review of intersecting brane
world models focusing on the main conceptional
ingredients. For more details we refer the reader to
the by now extensive literature.

In section 3 we derive the disc-level scalar potential (disc
tadpole) in the effective four dimensional theory arising from
intersecting brane worlds with D6-branes. Note that this potential
only depends on closed string moduli. The open string moduli
(positions of D-branes, Wilson lines) first appear in the open
string one-loop diagrams. The essential assumption of the whole
paper is that the ten-dimensional string coupling is small, so
that we can trust string perturbation theory. It is one of the
major unsolved problems in string theory by which actual
mechanism one can stabilize the  dilaton. Nevertheless,  in
all recent works on inflation from string theory this working
hypothesis has been made. On the contrary,
no assumption about a small curvature in the target space is required as
the potential we derive from divergences in one-loop
string amplitudes is exact to all orders in $\alpha'$. We will
introduce two different sets of coordinates for the
four-dimensional parameter space appearing in the open string tree
level potential. We call  the first set the ``gauge
coordinates", as they are proportional to the effective gauge
couplings on certain D-branes. Moreover, these coordinates would
naturally appear in an  ${\cal N}=1$ effective theory in four
dimensions. The second set is  called the ``Planck coordinates'',
for one of them is proportial to the four-dimensional Planck
scale.

In section 4 we will investigate whether this leading order
potential can satisfy the slow-rolling conditions
which are essential for an inflationary cosmological model.
The result can be summarized as follows. Working with the
``gauge coordinates'' we get a result very similar to \rqueb.
In fact the potential investigated there is just a very specific
example of the intersecting brane world potentials studied in this paper.
In fixing some of the ``gauge coordinates'', the remaining ones
can satisfy the slow-rolling condition. Such a mechanism was assumed
to consist in some, albeit unknown, dynamics that generates a potential
which then fixes the moduli.
However, not even two of them can be slow-rolling at the same time and
therefore all scalars except the inflaton would need to be fixed.
%Our interpretation of this result will turn out to be rather sceptical.
One generic feature when working with ``gauge coordinates'' is that
the string scale is evolving during the inflationary era as well.
The other option is to work with the ``Planck coordinates'',
assuming some most likely non-perturbative stringy mechanism to stabilize
the four-dimensional dilaton. This immediately implies
that both the Planck and the string scale are fixed.
In this case it turns out that the remaining three complex structure moduli
are generically {\it not} slow-rolling.

In section 5 we discuss a rather different scenario where
the complex structure
moduli are frozen at the string scale such that the relevant dynamics
affects the K\"ahler moduli exclusively.
A way to realize such a model in the context of intersecting brane worlds
consists in using orbifold background spaces \refs{\rbgka,\rbgkb,\rbgkc,\rfhs}
for compactifications of intersecting brane worlds of type
I string theory \refs{\rott,\rottb}. Another
option would be a model where the complex structure moduli are dynamically
stabilized by the
open string tree level scalar potential, which happens in type IIA
or type I intersecting brane worlds with some negative
wrapping numbers, i.e. with some effective anti-branes present.
The leading order contribution to the scalar potential then comes
from the one-loop
diagrams. By demanding some reasonable conditions to simplify the situation, the
only diagram relevant is the annulus, which not only depends on the closed
string K\"ahler moduli but also on open string fields, distances of the branes and
Wilson lines. We find that, interestingly,
the K\"ahler moduli are stabilized dynamically by the resulting potential,
leading
to small radii of the order of the string scale. In this regime
all the closed string moduli are frozen, such that the open string
moduli appear as inflaton candidates. While for large values
of the K\"ahler moduli slow-rolling in the open string fields is possible as
in \rquea, for small radii at the true minimum of the potential
this property is lost. Therefore, a cosmological application of this
scenario seems  unlikely.

\newsec{Toroidal Intersecting Brane Worlds}

The formalism for describing intersecting brane worlds in type I, type II
and even type 0$'$ string theory has been developed
in \refs{\rbgklnon\raads\rbgklmag\ras\rafiru\rafiruph\rbkl\rimr\rbonna\rott
\rcvetica\rcveticb\rbonnb-\berlin}.
Here, we mostly concentrate on type I strings but also include comments
on the modifications for the other models. In order to proceed, we
sketch the general idea and collect the formulae we need
in the remainder of the discussion.

The starting point is a T-dual version of type I string theory. In
toroidal models that we are going to use, one compactifies the
type IIA closed string on a six-dimensional torus $T^6$ which, for
simplicity, is taken to be of the form $T^6=\bigotimes_{I=1}^3
T_I^2$, each $T_I^2$ with coordinates $(X^I, Y^I)$ and radii
$R_1^I$ and $R_2^I$. Then the $\Omega {\cal R}$ orientifold
projection is performed with ${\cal R}$ being the reflection of
all three $Y^I$ coordinates. As a toroidal compactification of
type I theory down to four dimensions, this model has 16
supercharges and ${\cal N}=4$ supersymmetry in the closed string
sector. The orientifold projection introduces O6-planes stretching
along the fixed loci of ${\cal R}$, the $X^I$ directions, in
compact space and filling out the entire four-dimensional
uncompactified space-time.

In order to obtain the maximally symmetric solution that
corresponds to pure type I strings, one cancels the Ramond-Ramond
(RR) tadpoles induced in the non-orientable Klein bottle diagram
by placing D6-branes on top of the O6-planes. This leads to a
gauge group $SO(32)$ with 16 unbroken supercharges. However, the
RR tadpole cancellation conditions, stating that the sum of the
homological cycles of all the D6-branes is the same as the
homological cycle of the O6-planes \eqn\homo{
 \sum_{a=1}^K {N}_a \Pi_a=\Pi_{\rm O6} ,}
can also be satisfied  by
more general configurations of D6-branes which are no longer parallel but
intersect on the internal space \rbgklnon.
A D6$_a$-brane belonging to the stack $a\in\{ 1,...,K\}$ and wrapped on
a 3-cycle of the $T^6$ can be specified by the
wrapping numbers $(n_a^I,m_a^I)$ along the fundamental cycles of the torus.
Then \homo\ translates into conditions for these wrapping numbers and the
brane multiplicities $N_a$.
The effective low energy theory in four dimensions for such intersecting brane
models possesses interesting features which can be summarized as follows:

\item{$\bullet$} D-branes intersecting at angles generically break
         supersymmetry, so that at open string tree level supersymmetry
         is broken at the string scale.

\item{$\bullet$} There is a $U(N_a)$
                 gauge group on any stack of parallel D-branes.
In a supersymmetric vacuum a massless vectormultiplet of ${\cal
N}=4$ exists in the adjoint of the gauge group, which splits up
into massless and massive components when supersymmetry is broken.
Interactions will then generate potentials and masses for all
fields except for the gauge boson.

\item{$\bullet$} At an intersection point of two D6-branes there
                 appears a massless chiral fermion in the
bifundamental representation of the gauge groups $U(N_a)$ and
$U(N_b)$ on the two respective branes. Since two D-branes
                 on a torus may have multiple intersections, the
                 number of such fermions is degenerate and gives
                 rise to the number of families. The tadpole
                  cancellation conditions imply that the
                   effective theory is anomaly free.

\item{$\bullet$} The mass of the lowest scalar excitation in the open string
spectrum of strings stretching between two branes can be phrased in terms of their relative angles $\varphi_{ab}^I$ on the three tori (given here for
$\vert \varphi_{ab}^I \vert \le \pi/2$):
\eqn\masstach{
M^2_{\rm scal} = {1 \over 2\pi} \sum_{I=1}^3{| \varphi_{ab}^I |}
       - {1 \over \pi} {\rm max}\{ | \varphi_{ab}^I |\ :\ I= 1,2,3 \} .
}
Obviously, depending on the angles at the intersection,
                 a tachyonic scalar with negative mass can appear, which was
                 proposed to have the interpretation of a Higgs field
                 in the effective low energy theory \refs{\rbachas,\rbgklnon,\rafiruph}.
                 The phase transition signaled by this scalar may
                 simultaneously be responsible for a graceful exit from the
inflationary expansion of the early universe.
In fact, \masstach\ can be interpreted as a string realization of
a field dependent mass term as in \hybrd.

\meno
As mentioned, some of the above features are slightly modified in type II or
type 0$'$ theory. The general phenomenological patterns, however,
remain unaltered.
Motivated by these general perspectives, a systematic investigation
has revealed
that one can indeed find models with the Standard Model gauge group
and three families of appropriate chiral fermions. A more detailed study
of the phenomenological properties of these models was carried
out in \refs{\rafiru,\rafiruph,\rimr}.

\newsec{Tree level scalar potential}

Based on the earlier work \rquea\ where a simpler but less
realistic brane-anti-brane system was employed for modeling an
inflationary universe, the cosmological use of intersecting brane
world models was studied in \rrabadan. In both works, the dynamics
of the closed string moduli has been ignored and only the open
string scalars, the positions of the D6-branes on the internal
transverse space were considered. Only in the simplified brane-anti-brane
model this restriction recently has been relaxed \rqueb.
An argument in favor of such a simplification could be that in such
a brane-anti-brane setting the tree level tadpoles which drive the
dynamics of the transverse geometry are proportional to the
inverse of the volume of the space transverse to the branes.
In a large extra dimension scenario \refs{\radd,\raadd},
which may be advisable in order to avoid the
pitfall of the hierarchy problem anyway, this is large and the tadpoles
thus suppressed.
%gravitational effects may be weakened in the presence of large
%internal spaces transverse to the D-branes. Such a large extra
%dimension scenario  may be attractive also in order to avoid the
%pitfall of the hierarchy problem. In this approach the closed
%string tadpoles would be suppressed by ${\rm Vol}_\parallel / {\rm
%Vol}_\perp \ll 1$.
But this argument may turn out to be misleading, as the evolution
of the transverse volume under consideration can still be fast on
cosmological scales.

%since the
%internal volume is dynamical and may evolve much faster than the
%slowly rolling open string scalars under consideration, and thus the
%suppression could be spoiled soon.

In the following we will derive the leading order scalar potential
for the closed string moduli for general toroidal intersecting
D6-branes. This potential occurs already at open string tree level
and is exact to all orders in  $\alpha'$. It also represents the
potential for the untwisted moduli of the orbifold models considered
as another option for compactifications in section 5.

\subsec{The scalar potential in string frame}

In \rott\ the open string tree-level scalar potential for toroidal
intersecting brane worlds has been computed. The result can either
be extracted directly from the divergence in the one-loop
amplitudes or by integrating the Born-Infeld action for the
D6-branes
\eqn\dbi{ S_{\rm BI} = -T_p \int_{{\cal
M}_{p+1}}{d^{p+1}x\ e^{-\phi_{10}}\,
                          \sqrt{G_{p+1}}}}
over their compact world volume.
For type I intersecting brane worlds the potential is simply
\eqn\potena{ V=M_s^7\, e^{-\phi_{10}}\,
   \left( \sum_a N_a\,  V^{\rm D6}_a - V^{\rm O6}
                           \right) }
where $ V^{\rm D6}_a$ denotes the three-dimensional internal volume
of the D6$_a$-branes with wrapping numbers $(n_a^I,m_a^I)$
\eqn\potenb{ V^{\rm D6}_a=\prod_{I=1}^3 \sqrt{ \left( n_a^IR_1^I\right)^2+
                     \left( m_a^I R_2^I\right)^2  } }
and $ V^{\rm O6}$ denotes the internal volume of the O6-planes stretched along
the $X^I$ axes
\eqn\potenc{ V^{\rm O6}=-16 \prod_{I=1}^3 R_1^I .
}
For the case of intersecting D-branes in type IIA or even
in type 0$'$ string theory, the contribution from the orientifold
planes is absent. In type IIA this implies the absence of any net RR-charge
due to supersymmetry, but not so in type 0$'$. In this non-supersymmetric
string theory the orientifold planes are rather exotic objects that carry
charge but no tension.
Defining the complex structure moduli and the four-dimensional
dilaton as
\eqn\defi{  U^I={R_1^I\over R_2^I}, \quad
                     e^{-\phi_{4}}=M_s^3\, e^{-\phi_{10}}
                  \prod_I \sqrt{R_1^I\, R_2^I} ,}
one can express the disc level potential entirely in terms
of these variables
\eqn\potenh{ V_S(\phi_4,U^I)=   M_s^4\, e^{-\phi_4}\,
\left( \sum_{a=1}^K  N_a\, \prod_{I=1}^3
   \sqrt{\left( n_a^I\right)^2 U^I+
                          \left( m_a^I\, \right)^2
            {1\over U^I} }
          -16\, \prod_{I=1}^3 \sqrt{ U^I} \right) .}
The non-trivial
tadpole cancellation conditions phrased in terms of wrapping numbers read
\eqn\tad{
\sum_{a=1}^K{ N_a\, \prod_{I=1}^3{n^I_a} } -16 =
\sum_{a=1}^K{ N_a\, n_a^I m_a^J m_a^K } =0,
}
for any combination of $I\not= J \not= K \not= I$.
Furthermore, the brane spectrum is required to be invariant
under
\eqn\omegaproj{
\Omega {\cal R} : (n_a^I,m_a^I ) \mapsto (n_a^I, -m_a^I)
.}
Note that the potential only depends on the imaginary part $U^I$ of the
complex structure of the torus,
while its real part is frozen to take the value $b^I = 0$.
The four-dimensional dilaton and the complex structures $U^I$
appear to be the natural variables for expressing the
string frame leading order scalar potential. In the remainder of
this paper we call them ``Planck coordinates''.
Moreover, the potential for the imaginary part $T^I = M_s^2 R_1^I R_2^I$
of the K\"ahler structures
is flat at tree-level and thus can be neglected at this order.
We will come back to the K\"ahler moduli in section 5 when discussing
higher order one-loop corrections.

In ${\cal N}=1$ supersymmetric effective field theories
in four dimensions the particular combinations of scalars
\eqn\newk{\eqalign{  s&=M_s^3\, e^{-\phi_{10}}\, \prod_I R_1^I=
              e^{-\phi_{4}} \prod_I \sqrt{U^I} ,  \cr
                     u^I&= M_s^3\, e^{-\phi_{10}}\, R_1^I\, R_2^J\, R_2^K=
                   e^{-\phi_{4}}  \sqrt{U^I\over U^J\, U^K}\cr  }}
appear in chiral superfields such that the effective gauge couplings
can be expressed as a linear function of these variables  \rqsusy.
In terms of these ``gauge coordinates'' the
string frame scalar potential reads
\eqn\pot{\eqalign{
   V_S(s,u^I)= M_s^4\,&\sum_{a=1}^K  N_a\,
         \Biggl( \left(n_a^1 n_a^2 n_a^3\right)^2\, s^2+
                \sum_{I=1}^3 \left(n_a^I m_a^J m_a^K\right)^2\,
            \left( u^I \right)^2 \cr
    & + \left(m_a^1 m_a^2 m_a^3\right)^2 \left({u^1 u^2 u^3 \over s}\right) +
         \sum_{I=1}^3 \left(m_a^I n_a^J n_a^K\right)^2\,
               \left({s u^J u^K \over u^I}\right) \Biggr)^{1\over 2} -
        16\, M_s^4\,s, \cr}  }
where the last term  is the contribution from the O6-planes.

\subsec{The scalar potential in Einstein frame}

In order to discuss the slow rolling conditions
we have to transform to the Einstein frame
and need to make sure that the kinetic terms
for the scalar fields are canonically normalized.
%Thus, we have to transform the four-dimensional
%effective action into the Einstein frame.
In terms of the ``gauge coordinates'' the resulting
potential is given by
\eqn\potb{\eqalign{
   V_E(s,u^I) = M_{pl}^4 \, &\sum_{a=1}^K  N_a\,
              \biggl( \left(n_a^1 n_a^2 n_a^3\right)^2\, \left({1\over
                  u^1 u^2 u^3} \right)^2+
                \sum_{I=1}^3 \left(n_a^I m_a^J m_a^K\right)^2\,
            \left({1\over s\, u^J\, u^K}  \right)^2 + \cr
          &\left(m_a^1 m_a^2 m_a^3\right)^2
        \left({1 \over (s)^3\, u^1\, u^2\, u^3 }\right) +
         \sum_{I=1}^3 \left(m_a^I n_a^J n_a^K\right)^2\,
         \left({1 \over s\, (u^I)^3\, u^J\, u^K}\right) \biggr)^{1\over 2}\cr
         &-16\, M_{pl}^4\, \left({1\over u^1 u^2 u^3}\right) .} }
The rescaling to the Einstein frame performed above is simply defined by
\eqn\potc{   V_E(\phi_4,U^I)={M^4_{pl}\over M^4_s}\, e^{4\phi_{4}}
             \, V_S(\phi_4,U^I) .}
Since there is only one fundamental scale in string theory, one has the following
relation between the string scale $M_s$ and the Planck scale $M_{pl}$
\eqn\planck{   {M_s\over M_{pl}}= e^{\phi_4} = (s\, u^1\, u^2\,
    u^3)^{-{1/4}} . }
%\subsec{The kinetic terms}
Obviously, a running of any single one of the four fields $s,u^I$
at fixed $M_{pl}$ implies an evolution of the fundamental string scale
$M_s$.
After dimensional reduction to four dimensions, the kinetic
terms for the scalar fields read
\eqn\kinetic{ S_{kin}= M_{pl}^2 \int d^4 x\, \left[
           - (\partial^\mu \phi_4) (\partial_\mu \phi_4) -
           {1\over 4}\sum_{I=1}^3
           {(\partial^\mu \log U^I ) (\partial_\mu \log U^I)}\right] ,}
respectively
\eqn\kineticb{ S_{kin}= M_{pl}^2 \int d^4 x\, {1\over 4}\,\left[
            -(\partial^\mu \log s) (\partial_\mu \log s) -
           \sum_{I=1}^3
           {(\partial^\mu \log u^I ) (\partial_\mu \log u^I)}\right] .}
Thus, for the fields $s,u^I$ with a logarithmic derivative appearing in
\kinetic\ and \kineticb, the correctly normalized field are
$\tilde{s},\tilde{u}^I$ defined via
\eqn\norma{    {s}=e^{\sqrt 2 {\tilde{s}/M_{pl}}} , \quad
{u}^I=e^{\sqrt 2 {\tilde{u}^I/M_{pl}}} .
}
In the following section we will investigate these leading order
potentials in intersecting brane world models and look for
slowly rolling scalar fields.

\newsec{Complex structure and dilaton inflation}

Let us emphasize again, that our main assumption from the very
beginning is that we are allowed to work in string perturbation
theory. This means that the ten-dimensional string coupling has to
be small, i.e. $e^{\phi_{10}}\ll 1$. Next, since the open string
tree level potential is exact to all orders in $\alpha'$ we do {\it not} need
to impose that the internal radii are large compared to
the string scale, which is in contrast to \rqueb.
Moreover, we assume that all integer numbers
appearing in the intersecting brane world construction, like
the numbers $N_a$ of D-branes and the wrapping numbers $(n_a^I,m_a^I)$,
are not exceedingly
big. This assumption seems realistic, as with very big numbers
%we
%would have problems to satisfy the RR-tadpole cancellation
%conditions and
it would seem impossible to realize a reasonable
low energy particle spectrum.
%the three
%generation Standard Model of particle physics.

In order for one of the scalars $(\phi_4, U^I)$ or $(s,u^I)$ of
the closed string sector to be identified with the cosmological
inflaton field, its potential has to satisfy the two slow rolling
conditions: \eqn\slowroll{ \epsilon = {M_{pl}^2 \over 2} \left(
{V_E'(s,u^I) \over V_E(s,u^I)} \right)^2 \ll 1, \quad \eta =
M_{pl}^2 {V_E''(s,u^I) \over V_E(s,u^I)} \ll 1. } It is crucial
that the derivatives of $V(s,u^I)$ are taken with respect to
canonically normalized fields of \norma.

We will investigate these conditions under the assumption that three of
the relevant four scalar fields are frozen by some so far unknown mechanism,
and then check the slow-rolling condition for the remaining
scalar field. Depending on whether we work with the variables
$(s,u^I)$ or $(\phi_4,U^I)$, our conclusions will turn out to be
 very different.
Physically, the main difference between these two possibilities
is that, due to the relation \planck\ in the first case, the
string scale is forced to change during inflation,
whereas in the second case it can be made constant by freezing $\phi_4$.
In this sense the physical distinction between the respective
sets of coordinates
only plays a role when particular fields are frozen.
If one would not just consider to freeze some of the coordinate fields,
but also allow to impose relations among them, the distinction of coordinates
would turn irrelavant.

\subsec{Slow-rolling in ``gauge coordinates''}

As explained in \rqueb, the ``gauge coordinates''  $(s,u^I)$
are the natural coordinates to work with if we assume some
${\cal N}=1$ supersymmetric dynamics at some higher energy scale.

For toroidal type I intersecting brane world models, the breaking
of supersymmetry looks spontaneous in the sense that Str$({\cal
M}^2)=0$ in the open string spectrum \rbachas.
%In fact, the gauge theory sector
%on the D-branes has a spontaneously broken ${\cal N}=4$ supersymmetry.
But one needs to be careful in interpreting the breaking
mechanism. The potential \potb\ is in general not of the kind
which can occur as the scalar potential in a supersymmetric
theory. As long as no tachyon condensation in the open string
sector takes place, which would induce a discrete change of some
of the winding numbers $(n_a^I,m_a^I)$, while respecting charge
conservation, this property will be unaffected by the dynamics.
Hence, the theory is separated from a supersymmetric vacuum by a
phase transition. In particular, for a given set of winding
numbers, the theory will be non-supersymmetric at all scales.
There is actually a situation, where the potential indeed can be
cast into the form of a D-term potential in ${\cal N}=1$
supersymmetric theories which refers to adding an FI term in the
effective Lagrangian \rqsusy. At such a point, the potentially
tachyonic NS groundstate is just massless, the scalar signalling a
marginal deformation of the cycle wrapped by the two branes, which
is a continuous deformation of the theory on their world volume.
In this special situation, the non-supersymmetric effective theory
is in the same phase as the supersymmetric theory which appears
near the string scale.

Analogously to \rqueb, we will now investigate the slow-rolling conditions,
freezing three of the four ``gauge coordinates'' by hand.
Here, we will merely discuss the generic situation neglecting the logical
possibility that for
very specific choices of the wrapping numbers
or very special regions in parameter space  new features might appear.
\meno
$\bullet$ {\it $s$-Inflation}

Under the assumption that all three $u^I$ moduli are frozen, the
potential for the field $s$ without the contribution from the
orientifold planes is of the form \eqn\sinf{
V_E^{\rm D6}=M_{pl}^4 \sum_a N_a\, \left[
                \alpha_a+{\beta_a\over s^2} + {\gamma_a\over s}
                  +{\delta_a\over s^3} \right]^{1\over 2} ,}
where the coefficients can be read off from \potb\ and involve
the fixed scalars $u^I$ and some numbers of order one.
In particular, $\alpha_a > 0$. In the region $s\gg 1$ one has\footnote{$^1$}{For
the quasi-supersymmetric models discussed in \rqsusy,
all higher order corrections in $1/s$ are automatically absent
at string tree level.}
\eqn\sinfb{  V_E^{\rm D6}=M_{pl}^4 \left[ \left(\sum_a N_a\,
                \sqrt{\alpha_a} \right)+
                    {1\over 2} \left( \sum_a N_a\,
               {\gamma_a\over \sqrt{\alpha_a}}\right) {1\over s} +\ \cdots
\right] .} In type I the orientifold planes contribute \eqn\sinfc{
V_E^{\rm O6}=-M_{pl}^4\, {16\over \prod_I u^I} .} If we choose all
wrapping numbers $\prod_I n_a^I$ to be positive, then the constant
term in \sinfb\ cancels precisely against the O6-planes
contribution \sinfc\ due to the RR-tadpole cancellation
conditions. In this case, one simply gets $V\sim 1/s$, which
implies $V' \sim V$, with a constant of proportionality of order
one. Note that the derivative has to be taken with respect to the
canonically normalized field $\tilde{s}$, see \norma . Thus, this
case does not show slow-rolling behavior. However, if some of the
$\prod_I n_a^I$ are negative, which should generically be expected
to be the case, then we get a potential of the form \eqn\sinfd{
V_E= V_E^{\rm D6} + V_E^{\rm O6} = M_{pl}^4\left( A + B\,
           e^{-\sqrt 2{ \tilde{s}/M_{pl}}} + \ \cdots \right),}
using \norma. The above distinction actually applies to the case of type I
models, whereas in type II or type 0$'$ no negative orientifold contribution
appears in the potential, so that \sinfd\ always applies. Anyway, \sinfd\
is of the same kind as \sinfb\ in the absence of orientifold planes,
i.e. without \sinfc.
This potential is identical to the one which appeared in the recent analysis
in \rqueb. In fact, their configuration of D$9$- and D$5$-branes
is only a very specific choice of D9-branes with magnetic flux,
actually infinite
magnetic flux for a D5-brane, which
is just T-dual to the intersecting D6-branes studied in this paper.

For a potential of the form \sinfd\ the slow-rolling parameters
are readily computed to be
\eqn\slows{ \epsilon={B^2\over A^2}{1\over s^2}, \quad\quad
            \eta={2B\over A}{1\over s} ,}
such that $\eta\ll 1$ directly implies  $\epsilon\ll 1$.
Inserting the expressions for $\alpha_a$ and $\gamma_a$,
we deduce the following expression
\eqn\etan{\eqalign{ \eta=\sum_I \zeta_I {u^J\, u^K \over u^I\, s}
           =\sum_I {\zeta_I \over \left( U^I \right)^2} ,\cr}}
where the coefficients $\zeta_I$ are of order one. Thus, in order
to have slow-rolling, all complex structure moduli have to satisfy
$U^I\gg 1$ with their relative ratios fixed. Note that due to
\newk\ this is self-consistent with our assumption $s\gg 1$. We
have thus found that $s$ is a successful inflaton candidate under
the given assumptions. In the following we therefore discuss the
properties of $s$-inflation in some more detail.
\bigskip\bigskip
\bigskip

\meno
$\bullet$ {\it Exit from inflation}

Since both $A>0$ and $B>0$, we have $\eta > 0$ in \slows\ and
indeed face a positive cosmological constant. Hence $s$ rolls
towards larger values, i.e. deeper into the slow-rolling region.
Therefore, inflation can not end when the slow-rolling conditions
cease to be satisfied. However, intersecting D6-branes can have
open string tachyons localized on their intersection locus. Their
appearance depends on the angles of the intersecting branes, which
for fixed wrapping numbers depend on the  complex structure of the
internal torus, as was expressed in \masstach. Therefore, it
appears very suggestive that the entire system may evolve from a
tachyon free configuration of intersecting D6-branes, then during
inflation rolls down to a region in parameter space where tachyons
suddenly appear on certain intersections. These then dominate the
dynamics and will trigger a decay of the brane configuration to a
different, finally to a stable one. This solution to the graceful
exit problem is very reminiscent of the hybrid inflation scenario
\refs{\hybrid,\hybridb}. As the tachyon field localized at some
brane intersection carries the bifundamental representation of the
respective unitary gauge groups on the two stacks of branes, it is
well adapted to act as a Higgs field in the effective gauge
theory. Thus, the breakdown of inflation by the condensation of
D6-branes is interpreted as a phase transition that involves a
spontaneous breaking of gauge symmetry.

To investigate this possible exit in some more detail, we assume
that we have two intersecting D6-branes. At the starting point of
inflation, the three parameters for the angles under which they
intersect on the three $T_I^2$ are defined by
\eqn\intang{
\theta^I_0={1\over \pi}\, \arctan\left({\left({m^I_2\over n^I_2}+
          {m^I_1\over n^I_1}\right) U_0^I \over 1+\left({m^I_1\, m^I_2 \over
                n^I_1\, n^I_2}\right)\, \left(U_0^I\right)^2 }\right) ,}
indicating the initial values by their index 0. To simplify the
situation, they are supposed to satisfy $0<\theta^I<1/2$
throughout their evolution. The mass of the lowest bosonic mode is
given by \eqn\masstachb{ M^2_{\rm scal} = {1 \over 2} \sum_{I=1}^3
\theta^I
                  -  {\rm max}\{ \theta^I:I=1,2,3  \} . }
%begin figure
\fig{}{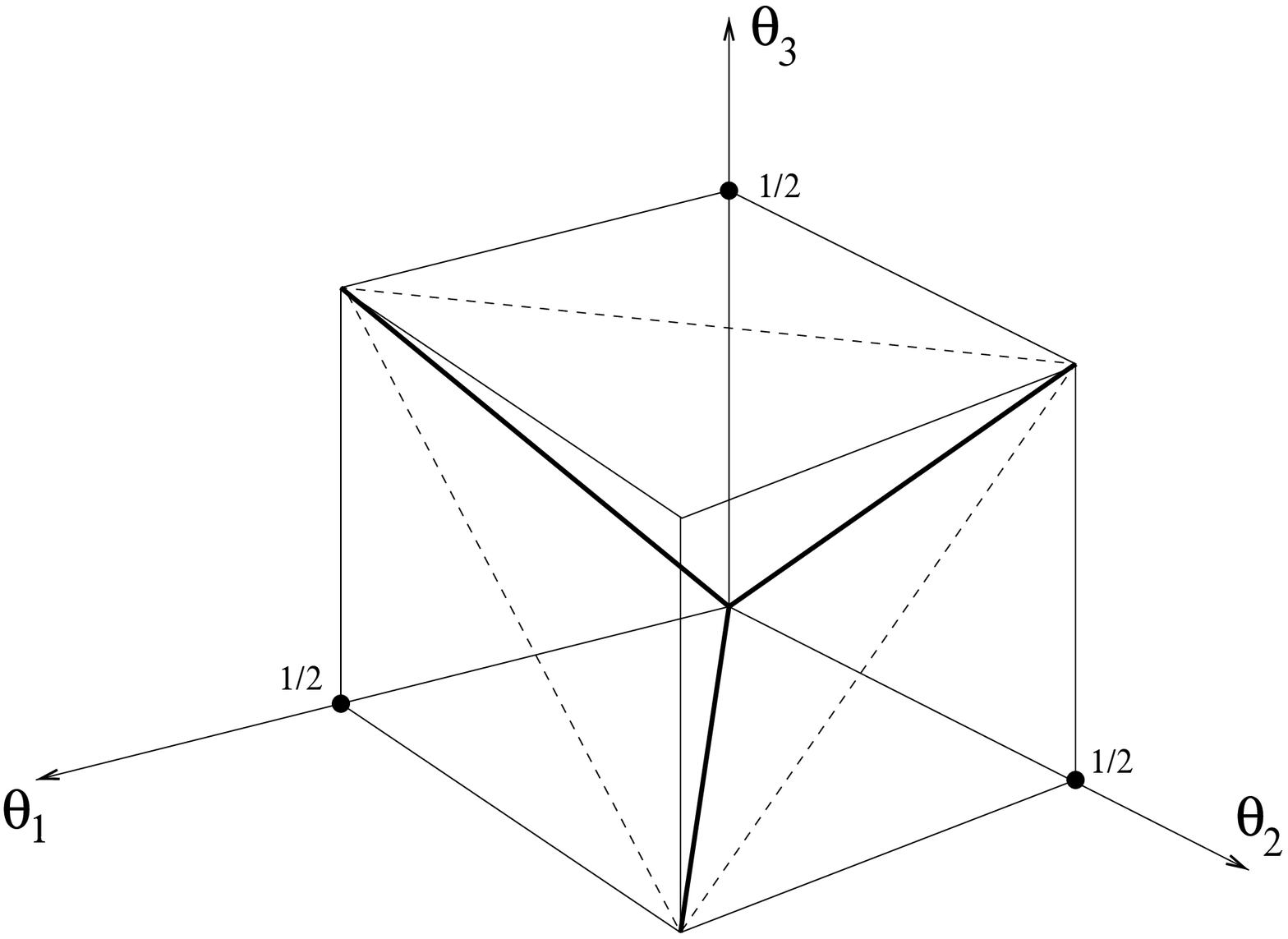}{14truecm}
%end figure
Thus, the region in $\theta^I$ space where no tachyons appear is
the interior of the cone with  edges
given by the bold diagonal lines in figure 1.
On  the faces of the cone the system preserves  ${\cal N}=1$
supersymmetry and on the edges ${\cal N}=2$ supersymmetry.
The origin corresponds to parallel branes which preserve
the maximal ${\cal N}=4$ supersymmetry.
During $s$-inflation the background geometry is driven towards larger values
of all three complex structures $U^I$ but with their ratios fixed.
Using \intang, we conclude that if none of the two D6-branes is parallel
to the $X^I$-axis, i.e. $m_a^I\not= 0$ for $a=1,2$,
then the intersection angle $\theta^I$ is driven to
$0$. If in fact any one of the two is parallel to the
$X^I$-axis, then $\theta^I$ goes to $1/2$.
To summarize, up to permutations we have the following four possible endpoints
of the flow
\eqn\endp{\eqalign{ (\theta^1,\theta^2,\theta^3)&\to (0,0,0),\quad\quad
                    {\cal N}=4\ {\rm SUSY,\ no\ tachyons} \cr
                    (\theta^1,\theta^2,\theta^3)&\to (0,0,1/2),\quad\quad
                    {\cal N}=0\ {\rm  SUSY,\ tachyons} \cr
                    (\theta^1,\theta^2,\theta^3)&\to (0,1/2,1/2),\quad\quad
                    {\cal N}=2\ {\rm SUSY,\ no\ tachyons} \cr
                    (\theta^1,\theta^2,\theta^3)&\to (1/2,1/2,1/2),\quad\quad
                    {\cal N}=0\ {\rm SUSY,\ no\ tachyons}. \cr}}
This classification actually leaves out the fact that the points
that the parameters are driven to cannot be reached within a given
set of winding numbers for any finite value of $U^I$. For instance
in the first case, two branes may approach vanishing intersection
angles very closely, but only if their $(n_a^I,m_a^I)$ were
proportional, they could become parallel. Thus, there may occur a
situation where a set of branes evolves towards an ${\cal N}=4$
supersymmetric setting dynamically, approaching it arbitrarily
well, but never reaching it without tachyon condensation. In fact,
tachyons can then no longer be excluded for such a brane setting
of the first type, as with three very small relative angles, the
mass of the NS groundstate may still become negative. But one
thing clearly can be deduced: Whenever the model contains two
intersecting D-branes, where one of the D-branes is parallel to
exactly one of the $X^I$-axes, the system evolves to a region
where tachyons do appear. Unfortunately,  it is difficult to
determine in general the precise end-point of inflation, i.e. the
point where the model crosses one of the faces in figure 1.

\meno
$\bullet$ {\it Number of e-foldings and spectrum of perturbations}

For an inflationary model to be successful it not only has to satisfy the
slow-rolling conditions, but also has to yield the right number of
e-foldings
\eqn\efold{ N=\int_{\tilde{s}_h}^{\tilde{s}_e} d\tilde{s} {1\over M_{pl}^2}\,
             {V_E\over V_E'}\simeq {A\over 2B}\,
             (s_{e}-s_{h})\simeq
          60-\log\left({10^{16}\ {\rm GeV} \over V_{\rm inf}^{1/4} }\right) ,}
where the index $h$ refers to horizon exit and the index $e$ to the
end point of inflation, while $V_{\rm inf}$ refers to the approximately
constant value of the potential $V_E$ during inflation.
Moreover, the amplitude of primordial density fluctuations \MukhanovXT\
in our case is
\eqn\density{ \delta_H\sim {1\over 5 \sqrt{3}\pi} \left({V_E^{3/2} \over
                      M^3_{pl}\, V_E'}\right) \simeq {1\over 5 \sqrt{6}\pi}
                       \left({A^{3/2} s_h \over B} \right), }
which must be tuned to yield the size of the observed temperature
fluctuations of the CMB $\delta_H=1.9\times 10^{-5}$. The spectral
index of the fluctuations is \eqn\spec{
n-1=-6\epsilon_h+2\eta_h\simeq 2\eta_h\simeq {4B\over A s_h} .} In
the present case of $s$-inflation we have $A,B>0$ so that $s_e\gg
s_h$, which implies $N=\eta_e^{-1}$. Without knowing $s_e$ and
$s_h$, we cannot make any more detailed prediction.

\bigno
$\bullet$ {\it $u^I$-Inflation}

If one freezes all ``gauge coordinates'' except one of the complex
structure moduli $u^I$, the story is very similar. Again, one gets
a potential of the form \sinfd\ in a $1/u^I$ expansion. Even
simpler, in this case the constant term $A$ never vanishes, not
even in type I models, so that the slow-rolling parameter can
always be written as \eqn\etam{\eta = \zeta_1 {u^J\, u^k \over
u^I\, s}
            +\zeta_2 {u^J\, s \over u^I\, u^K}
           +\zeta_3 {u^K\, s \over u^I\, u^J}
          ={\zeta_1\over  \left( U^I\right)^2} +\zeta_2
                         \left( U^J\right)^2 +\zeta_3 \left
                           ( U^K\right)^2, }
with the $\zeta_i$ of order one. Thus, slow rolling requires
$U^I\gg 1$ and $U^J,U^K\ll 1$ which is self-consistent with the
assumption $u^I\gg 1$. The constant $A$ is again positive but now
the constant $B$ can in principal become negative in type I, as
the orientifold  planes also contribute. In case it is negative,
the evolution would lead towards smaller values of $u^I$ until the
slow-rolling condition is no longer satisfied or open string
tachyons do appear. We have $u^I_h \gg u^I_e$ and, using
$u_h^I=-2\,B\,N/A$, we can express the density fluctuations in
terms of just $N$ and $A$ \eqn\den{ \delta_H \simeq {2\over 5
\sqrt{6}\pi}{A^{1/2}N},\quad\quad
            n-1=-{2\over N}. }
Note that by fixing only two of the four parameters, it is not
possible to satisfy the slow-rolling conditions for any two fields
among the $(s,u^I)$ at the same time. Thus, if we do not fix three
``gauge coordinates'', the potential is definitely not
slow-rolling. Fast rolling scalars then destabilize the background
before the slow-rolling of others becomes relevant cosmologically.
From the mathematical point of view, this result is almost
trivial. Qualitatively, in the ``Planck coordinates'' the
potential looks like a four-dimensional generalization of the
potential shown in figure 2. Starting at any point, there always
does exist a direction in which the potential does not change,
namely, if we move on a line of constant $V$. The only non-trivial
fact is that the directions along the ``gauge coordinates'' are
close to such lines of constant $V$ in their respective regions of
slow-rolling.
%begin figure
\fig{}{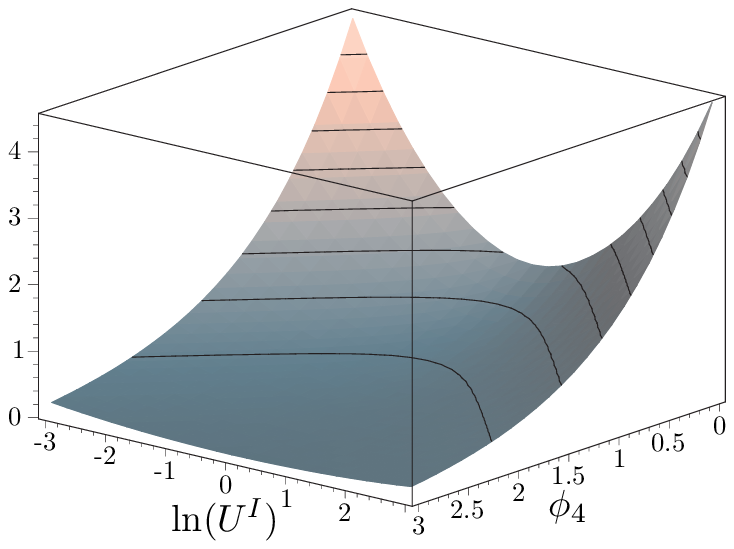}{14truecm}
%end figure
Given that the Planck scale is a fixed constant in nature,
$s$- and $u^I$-inflation implies that the string scale
must evolve during inflation. Thus, for $B$ being positive, the length of the string
inflates as well.

\subsec{Slow-rolling in ``Planck coordinates''}

In this section we investigate the slow-rolling properties, working
in the coordinates $(\phi_4,U^I)$.
If two D-branes intersect under non-trivial angles on all three $T_I^2$,
the annlus amplitude will only depend on the intersection angles and therefore
on the complex structure moduli.
These parameters are more natural in the string framework.
Since $\phi_4$ is apparently not
rolling slowly, we have to assume that this parameter is fixed by some
non-perturbative string dynamics. Note, that in this case
both the Planck-scale and the string scale are fixed, which seems
to be automatic from the string point of view.
\bigskip
\bigskip
\bigskip
\bigskip
\bigskip
\bigskip

\meno
$\bullet$ {\it $U^I$-inflation}

The only remaining candidate inflaton fields are  the complex
structure moduli $U^I$. The natural scale to freeze any of these
``Planck coordinates'' would in fact be the string scale. This can
either be achieved by imposing orbifold symmetries or by the tree
level potential itself. In \berlin\ we have shown that in type
0$'$ models such a dynamical stabilization takes place. Note that
the same happens im type II and type I intersecting brane world
models if some wrapping numbers $\prod_I n^I_a$ are chosen to be
negative. This option has been excluded in \rott\ implicitly,
where the run-away behaviour of the type I potential was
discussed. Assuming that we have fixed two of these complex
structures, the leading order potential for $U^I\gg 1$ is of the
form \eqn\leadpo{ V_E(U^I)=M_{pl}^4\, {A\, \sqrt{U^I}} . } Only in
type I with $\prod_I n^I_a>0$ for all $a$ one has \eqn\leadpob{
V_E(U^I)=M_{pl}^4\, {A \over \sqrt{U^I}} . } In any case, this
does not satisfy the slow-rolling conditions. A similar analysis
for $U^I\ll 1$ leads to the same negative result. The region in
the parameter space near to local or global minima of the
potential is less straight forward to discuss. Clearly, $\epsilon
\ll 1$ near an extremum, but usually $\eta \sim 1$ for all
examples we have studied, although we cannot present a general
prove for this statement. Thus, we conclude that in ``Planck
coordinates'' $(\phi_4,U^I)$ the disc level scalar potential is
not of an inflationary type upon freezing any of these
coordinates. \meno

\newsec{K\"ahler structure inflation}

The only option that we have in the ``Planck coordinates'' to
avoid the fast rolling of the complex structure moduli $U^I$ is to
freeze them. As it was shown in \rott, this can be achieved in
certain orbifold models from the very beginning while preserving
some of the intriguing phenomenological properties of intersecting
brane world models. Furthermore, for type 0$'$ backgrounds,
generically the complex structures are frozen dynamically at
values of order one, as it was shown in \berlin. The same holds
for type I models with some negative wrapping numbers $\prod_I
n^I_a$, as discussed above.

From now on, we assume that the complex structure moduli $U^I$ are
frozen in some way at order one and study the leading order
potential in the K\"ahler structure moduli $T^I=M_s^2 R_1^IR_2^I$.
The Einstein frame open string tree level potential now only
depends on the four-dimensional dilaton \eqn\tree{ V_E^{\rm
tree}=M_{pl}^4\, C\, e^{3\phi_4} ,} $C$ being a constant of order
one. A non-trivial dependence of the scalar potential on the
K\"ahler moduli can arise at the one-loop level. Assuming that the
closed string sector preserves supersymmetry, the torus and
Klein-bottle amplitude vanish. Moreover, the annulus and M\"obius
strip amplitude  can depend non-trivially on $T^I$ via possible
Kaluza-Klein (KK) and winding modes in non-supersymmetric sectors.
Therefore, the interesting sectors are those, where two D-branes
are parallel on one or two of the three $T_I^2$ in a way that this
open string sector still breaks supersymmetry. Beyond the K\"ahler
moduli, the KK and winding spectrum then also depends on open
string moduli, the distance $x$ between the branes and their
relative Wilson line $y$. Thus, the full potential is a function
of these three fields and the entire analysis will be more
involved. We will restrict to a rather simple case, where we
assume that the only contribution to the potential that depends on
the respective $T^I$ comes from a single set of two branes, which
are parallel there but not parallel to any of the O6-planes. The
only relevant amplitude left is then the annulus diagram of
strings stretching between the two branes. We further neglect any
dependence of the potential on $T^I$ other than $T^1=T$, which
finally is a function of only $(T,x,y)$. Up to this order in
perturbation theory it then explicitly takes the form
\eqn\oneloop{  V_E^{\rm 1-loop}(T,x,y)=M_{pl}\, M_s^3\, C_0  -
                        M_{s}^4\, C_1\, -
                        {\cal A}_{12}(T,x,y) .
}
The first two terms summarize contributions independent of $(T,x,y)$.
The situation described above can indeed be met in the intersecting
brane world orbifold
models of \rott.
The type 0$'$ models of \berlin\ do allow to stabilize the complex
structures at tree level,
but their bulk theory is not supersymmetric.

In this sense, we now consider the annulus amplitude for two
D-branes which are parallel on the first $T_1^2$ as the only
one-loop contribution to the potential that depends on $T$.
Then the open string KK and
winding spectrum can be written as
\eqn\kkw{   M^2_{op}=\Delta\left({(r+x)^2\over T} + T\, (s+y)^2 \right) ,}
where $\Delta$ is of order one and depends on the wrapping numbers
of the D-branes and the fixed complex structure $U^1$.
Here, $0\le y\le 1$ denotes the relative transversal
distance
between the two D-branes and $0\le x\le 1$ the relative Wilson line
along the longitudinal direction of the two D-branes on the $T_1^2$.
These variables are related to the canonical normalized open string
moduli by
\eqn\relati{   Y^2={1\over M_s^2}\, \Delta\, T\, y^2,\quad
               X^2={1\over M_s^2}\, {\Delta\over  T}\, x^2.}
The annulus loop channel amplitude reads
\eqn\anulus{\eqalign{
{\cal A}_{12}(T&,x,y)
={M_{pl}^4\over (8\pi^2)^2}\, e^{4\phi_4}\,
      N_1\, N_2\, I_{12}
                 \int_0^\infty {dt\over t^3}\,
              \left(\sum_{r,s\in\ZZ} e^{-2\pi t\, \Delta\, \left[
                 {(r+x)^2/T} + T (s+y)^2 \right]}\right) \cr
  &\times
 \left( {\th{0}{0}^2 \th{\epsilon_1}{0} \th{\epsilon_2}{0}-
         e^{-\pi i (\epsilon^1+\epsilon^2)}\,
                    \th{0}{{1\over 2}}^2 \th{\epsilon^1}{{1\over 2}}
                  \th{\epsilon_2}{{1\over 2}}-
                  \th{{1\over 2}}{0}^2 \th{\epsilon^1+{1\over 2}}{0}
             \th{\epsilon^2+{1\over 2}}{0} \over
                   \eta^6\, \th{\epsilon^1+{1\over 2}}{{1\over 2}}
                    \th{\epsilon^2+{1\over 2}}{{1\over 2}}
          e^{-\pi i (\epsilon^1+\epsilon^2+1)} } \right) ,\cr }}
where $I_{12}$ denotes the intersection number of the  two D-branes
on $T^4=T^2_2\times T^2_3$. The argument of the
$\vartheta$-functions is $q={\rm exp}(-2\pi t)$ and
$\epsilon^I=\varphi^I_{12}/\pi$
denotes the intersection angles of the two D-branes.
The NS ground state energy is (again take $0 < \epsilon^I < 1/2$)
\eqn\energ{  M^2_{\rm scal}=
     \left(\Delta {x^2\over T} + \Delta\, T\, y^2\right) +
                {1\over 2}\left(\epsilon^1+\epsilon^2\right)-{\rm max}
             \{\epsilon^I:I=1,2\} 
%-{1\over 2} 
}
which might be tachyonic depending on $(T,x,y)$.
Transforming the amplitude \anulus\ into tree channel via the
modular transformation $l=1/(2t)$ one obtains
\eqn\anulusb{\eqalign{
\widetilde{{\cal A}}_{12}(T,x,y)
=&{M_{pl}^4\over (8\pi^2)^2}\, e^{4\phi_4}\,
      N_1\, N_2\, I_{12}
                 \int_0^\infty {dl}\, 
     \left({1\over \Delta} \sum_{r,s\in\ZZ} e^{-\pi l\, \Delta^{-1}\, \left[
                 {T\, r^2} + {s^2/T} \right]}\,
              e^{-2\pi i (r\,x+s\, y)}\right) \cr
  &\times   \left( {\th{0}{0}^2 \th{0}{\epsilon^1} \th{0}{\epsilon^2}-
                   \th{0}{{1\over 2}}^2 \th{0}{\epsilon^1+{1\over 2}}
             \th{0}{\epsilon^2+{1\over 2}}-
                    \th{{1\over 2}}{0}^2 \th{{1\over 2}}{\epsilon_1}
                  \th{{1\over 2}}{\epsilon^2} \over
                   \eta^6\, \th{{1\over 2}}{\epsilon^1+{1\over 2}}
                    \th{{1\over 2}}{\epsilon^2+{1\over 2}}  } \right) ,\cr }}
where the argument of the $\vartheta$-functions is
$\tilde q={\rm exp}(-4\pi l)$.
Since all the non-supersymmetric vacua suffer from the presence of a
NSNS tadpole, the one-loop amplitude
contains divergences coming from the exchange of massless
modes between the two D-branes.
Thus, to continue, we have to regularize the expresssion
\anulusb\ by substracting the divergent piece
\eqn\regu{  \widetilde{{\cal A}}_{12}^{\rm reg}(T,x,y)
=\widetilde{{\cal A}}_{12}(T,x,y)
                          - \widetilde{K}_{12} }
with
\eqn\regub{  \widetilde{K}_{12}={M_{s}^4\over (8\pi^2)^2}\,
      N_1\, N_2\, I_{12}
                 \int_0^\infty {dl}\, {4\over \Delta}\,
             {\sin^2\left({\pi(\epsilon_1+\epsilon_2) \over 2}\right)\,
              \sin^2\left({\pi(\epsilon_1-\epsilon_2) \over 2}\right) \over
                \sin(\pi \epsilon_1)\,  \sin(\pi \epsilon_2) } .}
Concerning the dynamics of $(T,x,y)$ the subtraction appears to be
unimportant, since it does not depend on these fields.
The first point to notice is that
the potential \oneloop\ indeed stabilizes the K\"ahler modulus $T$
dynamically, which can be seen as follows.
 For $x=y$ the KK and winding sum is invariant under
a T-duality, which maps $T\to 1/T$ and exchanges
$x\leftrightarrow y$. Thus, there is an extremum at the
self-dual point $T=1$, which fixes the internal radii at values of the order
of the string scale. We have numerically evaluated the regularized
annulus amplitude for specific choices of the angles and
confirmed this expectation.

With $T$ thus frozen, the open string modulus $x=y$ could
be a candidate inflaton field,
if it satisfies the slow-rolling conditions, assuming that $x=y$
is in fact dynamically stable.
For $T\gg 1$ and fixed a very similar result
 has in fact been proven in \rquea\ in the neighbourhood
of the instable antipodal  point. The essential observation  there was
that the second derivate $V''$ of the potential at the antipodal  point
vanishes, so that not only the slow-rolling paramter $\epsilon$ but
also $\eta$ becomes arbitarily small.
The question we now want to address is, whether this behavior still exists, if
$T$ comes close to its true minimum value at $T=1$. Then, not only
massless modes contribute to the force between the two D6-branes
but also massive string excitations.

For analyzing this question we expand the contribution to \anulusb\
from the $\vartheta$-functions in a $\tilde q$-series.
Considering first  the $\tilde q^0$ term and summing over all
KK and winding modes, one essentially has to evaluate the integral
\eqn\inti{ \int_0^\infty dl \left[-1+\left( 1+2\sum_{r\ge 1}
                  e^{-\pi l\, \Delta^{-1}\,
                 {T\, r^2} } \cos (2\pi  r x) \right)
                 \left( 1+2\sum_{s\ge 1}
                  e^{-\pi l\, \Delta^{-1}\,
                 { {s^2\over T}} } \cos (2\pi s y) \right)\right] .}
This can be done straighforwardly. By expanding the result around the
``symmetric antipodal'' point, using $x=1/2-\o x$, $y=1/2-\o y$,
we find that the
linear and the quadratic terms in the fluctuations
$\o x$ and $\o y$ precisely vanish.
This computation exactly yields the large distance result
of \rquea.
However, the minimum for $T$ is not at large distances,
but at distances of the order the string scale.
Now, by taking the $\tilde q^1$ term into acount and performing the
same computation, we find that still the linear terms in
 $\o x,\ \o y$ vanish, but that the quadratic terms do not.
Thus, taking the exchange of massive string modes into account
destroys the slow-rolling property $\eta \ll 1$. We have also done a numerical
analysis and found this result confirmed.

Summarizing, in intersecting brane world models
the leading order potential for the K\"ahler moduli
is well suited to stabilize these dynamically.
But at the real minimum of its potential near $T=1$ the well appreciated
slow-rolling properties of the open string moduli for large $T$ are lost.
A cosmological application therefore appears problematic.

\newsec{Conclusions}

In this paper we have analyzed intersecting brane world models
with respect to their ability to give rise to inflation. Our starting point
was the open string tree level scalar potential transformed
to the Einstein frame. We have investigated this potential for two different
scenarios. In the first scenario, as proposed in \rqueb,
we have assumed that some of the ``gauge coordinates'' are frozen
by an unknown mechanism.
Analogously to \rqueb\ we have found that the potential generically satisfies
the slow-rolling conditions both for $s$-inflation and for $u^I$-inflation,
if the remaining three moduli are frozen by hand.

In the second scenario we have worked with the ``Planck coordinates''.
Even after fixing the four-dimensional dilaton by hand, the
potential for the complex structures was generically not of slow-rolling type.
On the one hand it might look encouraging to find inflation at least
in one scenario, on the other hand it is of course disappointing
that one has to freeze rather artificially a very particular choice
of scalar fields, without knowing any explicit mechanism which could do
this job.

Finally, we also have studied the leading order potential for the
K\"ahler and open string
moduli in models where the running of the complex structure
moduli was frozen. First of all, we
found that the K\"ahler moduli are generically stabilized dynamically.
However, again rather disappointingly, the slow-rolling
properties of the open string moduli at antipodal points, which were
appreciated in simpler models, get lost for small values of the internal radii
at the minimum of the scalar potential.

Apparently, we are still far away from a viable and realistic
string theoretic realization of inflation. For all scalar
potentials coming from string theory that have been studied so far,
one can only achieve inflationary models by freezing some of the
moduli by hand. At this time, this seems to be the state of
the art and any substantial improvement would have to be considered a
real progress in relating string theory to inflationary cosmology.

%\appendix{A}{Derivation of the moduli kinetic terms}
%The lagrangian for the kinetic ten dimensional dilaton and the
%Einstein-Hilbert terms can be written as \eqn\lagn{ M_s^8 e^{ -
%2\phi _{10} } \sqrt {{\rm  - G}} {\rm G}^{MN} \left(
%{{\textstyle{1 \over 2}}{\rm R}_{MN}  + 2\partial _M \phi _{10}
%\partial _N \phi _{10} } \right).}
%The metric for the used toroidal compactification takes the form
%\eqn\metric{{\rm G}^{MN}=diag\left( {g_{\mu \nu } (x),r_1 ^2
%(x),r_2 ^2 (x),r_1 ^2 (x),r_1 ^2 (x),r_1 ^2 (x),r_1 ^2 (x)}
%\right)}
%with $g_{\mu \nu }$ being the four dimensional uncompactified metric.
%{\bf ...Rest kommt noch.}

\bigskip
\centerline{{\bf Acknowledgments}}\pano
B.~K. wants to thank Michael Haack for helpful discussions.
The work is supported in part by the European Community's Human Potential
Programme under contract HPRN-CT-2000-00131 Quantum Spacetime.
T.~O. also likes to thank the Graduiertenkolleg {\it The Standard
Model of Particle Physics - structure, precision tests and extensions},
maintained by the DFG.

\vfill\eject

\listrefs

\bye
\end